\documentclass[prc,reprint,twoside,showpacs,amsmath]{revtex4-1}

\usepackage{graphicx}
\usepackage[colorlinks=true,allcolors=blue]{hyperref}

\newcommand\bra\langle
\newcommand\ket\rangle
\newcommand\GL{\text{GL}}
\newcommand\SL{\text{SL}}
\newcommand\SO{\text{SO}}
\newcommand\Sp{\text{Sp}}
\newcommand\USp{\text{USp}}
\newcommand\mb{\bar m}
\newcommand\alb{\bar\alpha}
\newcommand\vb{\bar v}
\newcommand\cV{\mathcal V}
\newcommand\cS{\mathcal S}
\newcommand\vS{\mathbf S}
\newcommand\vT{\mathbf T}
\newcommand\vP{\mathbf P}

\begin{document}

\title{Conservation laws in the $1f_{7/2}$ shell model of $^{48}$Cr}

\author{K. Neerg\aa rd}

\affiliation{Fjordtoften 17, 4700 N\ae stved, Denmark}

\begin{abstract}
  Conservation laws in the $1f_{7/2}$ shell model of $^{48}$Cr found
  in numeric studies by Escuderos, Zamick, and Bayman [A. Escuderos,
  L. Zamick, and B. F. Bayman, arXiv:nucl-th/0506050 (2005)] and me
  [K. Neerg\aa rd, Phys. Rev. C \textbf{90}, 014318 (2014)] are
  explained by symmetry under particle-hole conjugation and the
  structure of the irreps of the symplectic group Sp(4). A
  generalization is discussed.
\end{abstract}


\maketitle

\section{\label{sec:intr}Introduction}

In a recent article, I analyzed the ground states of the nuclei
$^{48}$Cr, $^{88}$Ru, and $^{92}$Pa calculated in the $1f_{7/2}$ or
$1g_{9/2}$ shell model with effective interactions from the literature
in terms of irreps of the symplectic group $\Sp(2j+1)$, where $j$ is
the single-nucleon angular momentum~\cite{Ne14}. A surprising
observation emerged from this study: In $^{48}$Cr, independently of
the interaction, the interaction matrix elements between the irrep
$vt=61$ and other irreps vanishes within the numeric accuracy. Here,
$v$ denotes the seniority, and $t$ the reduced isospin~\cite{Fl52},
and I have introduced a notation to be used in this article: For
example, $vt=61$ is shorthand for $(v,t)=(6,1)$. In~\cite{Ne14}, I had
to leave this observation unexplained; it is explained below.

The phenomenon turns out to be related to another one observed in the
literature, by Escuderos, Zamick, and Bayman~\cite{Es05}. In
calculations of the entire spectrum of $^{48}$Cr in the $1f_{7/2}$
shell model with a certain interaction, these authors found that
states with opposite parities of $(v_n+v_p)/2$, where $v_n$ and $v_p$
are the seniorities~\cite{Ra43,Ra49} of the neutron and proton
systems, did not mix within the numeric accuracy. I was informed in a
private communication with one of them that they are aware of the
origin of this conservation of the parity of $(v_n+v_p)/2$ in a
midshell nucleus in the symmetry of the model under particle-hole
conjugation~\cite{Za14}.

The conservation of the $\Sp(2j+1)=\Sp(8)$ irrep $vt=61$ in $^{48}$Cr
for angular momentum and isospin $I=T=0$ and several similar laws of
conservation of $\Sp(2j+1)$ irreps in midshell nuclei will be shown to
follow from this conservation law. The connection is established by
means of the Helmers duality~\cite{He60} and the structure of the
irreps of the group $\Sp(4)$. These matters are discussed in
Sec.~\ref{sec:class}. Then in Sec.~\ref{sec:conj}, I review the theory
of particle-hole conjugation and derive formally the conservation of
the parity of $(v_n+v_p)/2$ in a midshell nucleus. In
Sec.~\ref{sec:vt}, these pieces are assembled into an explanation of
the aforesaid observation in~\cite{Ne14}, and this observation is
generalized. The article is summarized in Sec.~\ref{sec:sum}.

\section{\label{sec:class}Classification of states in a $j$ shell}

Helmers has shown that the configuration space $\cV$ of the system of
all possible numbers of fermions of different kinds $\tau=1,\dots,k$
in a $j$ shell, or more generally, in any single-fermion space of even
dimension $2\Omega=2j+1$, is composed of different single irreps of
$\Sp(2\Omega)\times\Sp(2k)$, where $\Sp(2d)$ is the symplectic group
in $2d$ dimensions~\cite{He60}. [Helmers considers, equivalently, the
unitary subgroup $\USp(2\Omega)\times\USp(2k)$.] The irreps of
$\Sp(2d)$ are associated with Young frames with at most $d$
rows~\cite{We39}. In the decomposition of $\cV$, the irreps of
$\Sp(2\Omega)$ have Young frames with at most $k$ columns, and the
irreps of $\Sp(2k)$ have Young frames with at most $\Omega$ rows, and
they are combined so that the lengths of the $i$th row in the
$\Sp(2\Omega)$ frame and the $(\Omega+1-i)$th column in the $\Sp(2k)$
frame add up to $k$. Equivalently, the lengths of the $i$th column in
the $\Sp(2\Omega)$ frame and the $(k+1-i)$th row in the $\Sp(2k)$
frame add up to $\Omega$.

Due to the even dimension of the single-fermion space, its basic
states can be combined in $\Omega$ pairs, which I label $(m,\mb)$. In
a $j$ shell, $m$ can be taken as the magnetic quantum number
restricted to $m>0$, and $\mb$ may label the time reversed state so
that with usual phase conventions~\cite{Ed57},
$|\mb\ket=(-)^{j+m}|\!-\!m\ket$. Helmers pairs $|m\ket$ with
$(-)^{j-m}|\!-\!m\ket=-|\mb\ket$. The present choice will prove its
advantage in later considerations of time reversal. A basis for the
representation of the $\Sp(2\Omega)$ infinitesimal algebra is formed by
the operators
\begin{equation}\label{eq:genSp2Om}\begin{gathered}
    A_{mm'} = \sum_{\tau} (a_{m\tau}^\dag a_{m'\tau}
    - a_{\mb'\tau}^\dag a_{\mb\tau}), \\
    B_{mm'} = B_{m'm} = \sum_{\tau} (a_{\mb'\tau}^\dag a_{m\tau}
    + a_{\mb\tau}^\dag a_{m'\tau}), \\
    B_{mm'}^\dag = B_{m'm}^\dag,
\end{gathered}\end{equation}
where $a_{m\tau}$ and $a_{\mb\tau}$ annihilate a fermion in the states
$|m\tau\ket$ and $|\mb\tau\ket$. A basis for the representation of the
$\Sp(2k)$ infinitesimal algebra is formed by the operators
\begin{equation}\label{eq:genSp2k}\begin{gathered}
  \mathcal N_{\tau\tau'} = \sum_m (a_{m\tau}^\dag a_{m\tau'}
    - a_{\mb\tau'} a_{\mb\tau}^\dag), \\
  P_{\tau\tau'} = P_{\tau'\tau} = \sum_m (a_{\mb\tau'} a_{m\tau}
    + a_{\mb\tau} a_{m\tau'}), \\
  P_{\tau\tau'}^\dag = P_{\tau'\tau}^\dag.
\end{gathered}\end{equation}
Notice $\mathcal N_{\tau\tau}=n_\tau-\Omega$, where $n_\tau$ is the
number of fermions of kind $\tau$. It is easily verified that each
operator \eqref{eq:genSp2Om} commutes with each operator
\eqref{eq:genSp2k}.

The best known case is that of identical fermions, $k=1$. There the
$\Sp(2\Omega)$ Young frame has only one column, whose length is the
seniority $v$~\cite{Ra43,Ra49}. The infinitesimal algebra of
$\Sp(2k)=\Sp(2)$ is isomorphic to the angular momentum algebra. This
is Kerman's quasispin algebra~\cite{Ke61} with basic operators
\begin{equation}\label{eq:quspin}
  S^0 = \tfrac12 (n-\Omega), \quad
  S^- = \tfrac12 P, \quad
  S^+ = \tfrac12 P^\dag,
\end{equation}
where the index $\tau=1$ is omitted for convenience. With
$\vS^2=\cS(\cS+1)$, where $\vS$ is the vector with complex coordinates
\eqref{eq:quspin}, the length of the single row of the $\Sp(2)$ Young
frame is $2\cS$. The general rule for the correspondence of the
$\Sp(2\Omega)$ and $\Sp(2k)$ Young frames gives the well known
relation~\cite{Ta93}
\begin{equation}\label{eq:Sv}
  \cS = \tfrac12 (\Omega-v).
\end{equation}

For $k>1$, we have
\begin{equation}\label{eq:Vdec}
  \cV = \bigotimes_\tau \cV_\tau,
\end{equation}
where $\cV_\tau$ is the space of states where only fermions of kind
$\tau$ are present. Acting within each such space we have groups
$\Sp(2\Omega)_\tau$ and $\Sp(2)_\tau$ with infinitesimal algebras
spanned by
\begin{equation}\label{eq:genSp2Omtau}\begin{gathered}
  A_{\tau,mm'} = a_{m\tau}^\dag a_{m'\tau} - a_{\mb'\tau}^\dag a_{\mb\tau}, \\
  B_{\tau,mm'} = B_{\tau,m'm} = a_{\mb'\tau}^\dag a_{m\tau}
    + a_{\mb\tau}^\dag a_{m'\tau}, \\
  B_{\tau,mm'}^\dag = B_{\tau,m'm}^\dag,
\end{gathered}\end{equation}
and
\begin{equation}\label{eq:quspintau}
  S^0_\tau = \tfrac12 (n_\tau-\Omega), \quad
  S^-_\tau = \tfrac12 P_{\tau\tau}, \quad
  S^+_\tau = \tfrac12 P^\dag_{\tau\tau}.
\end{equation}
Their irreps can be labeled by seniorities $v_\tau$ and quasispins
$\cS_\tau$, respectively, related by \eqref{eq:Sv} separately for each
$\tau$, that is,
\begin{equation}\label{eq:Svtau}
  \cS_\tau = \tfrac12 (\Omega-v_\tau).
\end{equation}
Any pair of operators \eqref{eq:genSp2Omtau} and \eqref{eq:quspintau}
with different $\tau$ evidently commute.

Particularly relevant for nuclear physics is the case $k=2$ with the
two kinds of fermions being neutrons and proton. I denote in this case
the values of $\tau$ accordingly by $n$ and $p$. Its $\Sp(2\Omega)$
Young frames have at most two columns and give rise to Flowers's
classification of the states in $\cV$ with given number of nucleons
and isospin by a seniority $v$ and a reduced isospin $t$~\cite{Fl52}.
This seniority $v$ is the total number of cells of the Young frame,
and $2t$ is the difference of its column lengths. The infinitesimal
algebra of $\Sp(2k)=\Sp(4)$ is isomorphic to that of the proper
orthogonal group $\SO(5)$ and may be better known by this
name~\cite{En96}. Its dimension is ten. The four basic operators
$\mathcal N_{\tau\tau'}$ combine to $n-2\Omega$, where $n=n_n+n_p$ is
the number of nucleons, and coordinates of the isospin $\vT$, the
three linearly independent basic operators $P_{\tau\tau'}$ to
coordinates of a pair annihilation isovector $\vP$, and the three
linearly independent basic operators $P_{\tau\tau'}^\dag$ to
coordinates of a pair creation isovector $\vP^\dag$.

\begin{figure}
  {\center\includegraphics[width=\columnwidth]{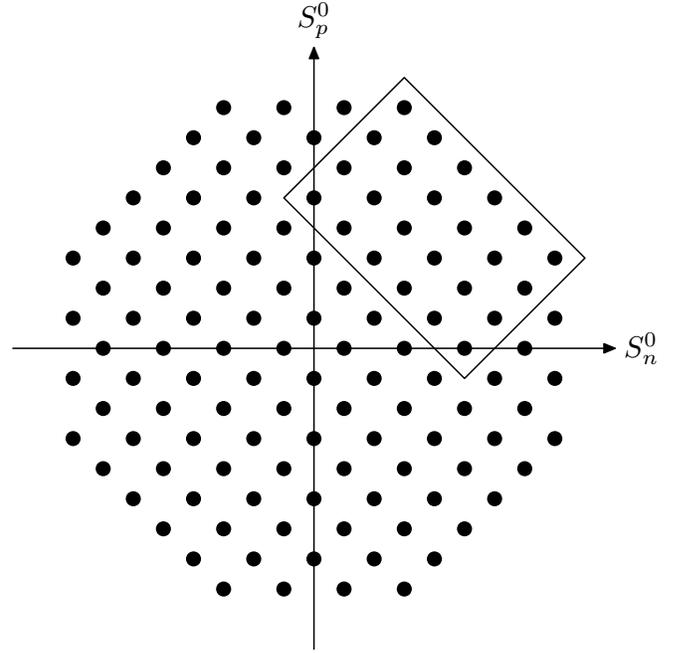}\par}
  \caption{\label{fig:genw}Weighs of the $\Sp(4)$ irrep
    $[\lambda\mu]=[83]$. Horizontal or vertical neighbors have
    distances one. The rectangle is the border of the range
    \eqref{eq:Sp4>Sp2nxSp2p}.}
\end{figure}

I label the $\Sp(4)$ irreps in the conventional notation for Young
frames (or partitions) by $[\lambda\mu]$, where
\begin{equation}\label{eq:lammu}
  \left.\begin{aligned} \lambda & \\ \mu & \end{aligned}
  \right\rbrace = \Omega - \tfrac12 v \pm t
\end{equation}
are the row lengths of the Young frame. The operators $S^0_\tau$ form
a basis for a Cartan subalgebra of the infinitesimal algebra. Adopting
this basis I refer to the eigenvalues of $(S^0_n,S^0_p)$ that occur in
a given representation as the \textit{weights} of this representation.
For the irrep $[\lambda\mu]$ they obey~\cite{Ra51}
\begin{equation}\label{eq:weights}\begin{gathered}
  \tfrac12(\lambda + \mu) + S^0_n + S^0_p \text{ is integral}, \\
  |S^0_\tau| \le \tfrac12\lambda\ \text{for}\ \tau = n,p, \\
  |S^0_n \pm S^0_p| \le \tfrac12(\lambda + \mu).
\end{gathered}\end{equation}
Figure~\ref{fig:genw}
shows this range in a generic case.

Because $\Sp(2)_n$ commutes with $\Sp(2)_p$, the $\Sp(4)$ irrep
$[\lambda\mu]$ is the direct sum of irreps of
$\Sp(2)_n\times\Sp(2)_p$. I shall show that this direct sum includes
exactly once each tensor product of $\Sp(2)_n$ and $\Sp(2)_p$ irreps
with quasispins $\cS_n$ and $\cS_p$ such that $(\cS_n,\cS_p)$ is a
weight, and
\begin{equation}\label{eq:Sp4>Sp2nxSp2p}
  |\cS_n - \cS_p| \le t \le \cS_n + \cS_p.
\end{equation}
The border of this range is shown as a rectangle in
Fig.~\ref{fig:genw}. It follows that a basis for the $\Sp(4)$ irrep is
labeled uniquely by the quantum numbers $\cS_n$, $\cS_p$, $S^0_n$, and
$S^0_p$. Each such basic vector is associated with an irrep of
$\Sp(2\Omega)$ in $\cV$ whose states have the quantum numbers $v$ and
$t$ corresponding by \eqref{eq:lammu} to $\lambda$ and $\mu$, as well
as the quantum numbers $v_n$, $v_p$, $n_n$, and $n_p$ corresponding by
\eqref{eq:Svtau} and \eqref{eq:quspintau} to $\cS_n$, $\cS_p$,
$S^0_n$, and $S^0_p$.

To prove the assertion, I consider characters. Weyl derived a general
formula for the characters of $\Sp(2d)$~\cite{We25}. We need the
$\Sp(2)$ characters
\begin{equation}\label{eq:Sp2ch}
  \chi_{[\lambda]}(u)=\frac{u^{\lambda+1}-u^{-\lambda-1}}{u-u^{-1}},
\end{equation}
where $u$ and $u^{-1}$ are the eigenvalues of the symplectic
transformation, and the $\Sp(4)$ characters
\begin{equation}\label{eq:Sp4ch}
  \chi_{[\lambda\mu]}(u,v)=\frac
    {\left|\begin{array}{cc}
       u^{\lambda+2}-u^{-\lambda-2} & u^{\mu+1}-u^{-\mu-1} \\
       v^{\lambda+2}-v^{-\lambda-2} & v^{\mu+1}-v^{-\mu-1}
     \end{array}\right|}
    {\left|\begin{array}{cc}
       u^2-u^{-2} & u-u^{-1} \\
       v^2-v^{-2} & v-v^{-1}
     \end{array}\right|},
\end{equation}
where $u$, $v$, $u^{-1}$, and $v^{-1}$ are the eigenvalues of the
symplectic transformation. Because each matrix in
$\Sp(2)_n\times\Sp(2)_p$ as a subgroup of $\Sp(4)$ is the direct sum
of the matrices from $\Sp(2)_n$ and $\Sp(2)_p$, what must be proven is
\begin{equation}\label{eq:chrel}
  \chi_{[\lambda\mu]}(u,v) = \sum_{\cS_n\cS_p}
    \chi_{[2\cS_n]}(u) \, \chi_{[2\cS_p]}(v),
\end{equation}
with summation over $[\lambda\mu]$ weights $(\cS_n,\cS_p)$ obeying
\eqref{eq:Sp4>Sp2nxSp2p}. It takes a straightforward calculation to
verify this algebraic identity.

\section{\label{sec:conj}Particle-hole conjugation}

Condon and Shortley~\cite{Co35} and Racah~\cite{Ra42} introduced the
concept of particle-hole conjugation in a shell in a context of atomic
physics. It is noticed already in this early work that symmetry under
particle-hole conjugation gives rise to special conservation laws in
systems where the shell is half filled. I offer a derivation which is
more general and seems simpler than ones I have found in the
literature and clearly displays the few required assumptions.

In a $j$ shell, particle-hole conjugation may be defined~\cite{Be59}
as a unitary transformation determined within an unimportant phase
factor by
\begin{equation}\label{eq:conj}
  a_{m\tau} \mapsto a^\dag_{\mb\tau}, \quad 
  a_{\mb\tau} \mapsto -a^\dag_{m\tau}
\end{equation}
for any $m$ and $\tau$. This implies
\begin{equation}\label{eq:conjn}
  n_\tau \mapsto 2\Omega - n_\tau.
\end{equation}
For the nuclear system with $k=2$, the complex coordinates $T^\kappa$
of the isospin vector $\vT$ transform by
\begin{equation}\label{eq:conjT}
  T^0 \mapsto -T^0, \quad  T^\pm \mapsto T^\mp,
\end{equation}
so that
\begin{equation}\label{eq:conjT^2}
  \vT^2 \mapsto \vT^2,
\end{equation}
that is, particle-hole conjugation commutes with $T$.

The transformations \eqref{eq:conj} are easily seen to be realized by
the transformation of any operator $X$ by
\begin{equation}\label{eq:trans}
  X \mapsto U^\dag X U
\end{equation}
with
\begin{equation}\label{eq:U}
  U = \exp \frac\pi 2 (S^+\!-\!S^-) = \exp i \pi S^y 
\end{equation}
in terms of coordinates $S^\kappa$ of
\begin{equation}\label{eq:S}
  \vS = \sum_\tau \vS_\tau,
\end{equation}
where $\vS_\tau$ is the vector with the complex coordinates
\eqref{eq:quspintau}. The transformation \eqref{eq:U} is the rotation
by the angle $-\pi$ about the $y$ axis in quasispin space,
so~\cite{Ed57}
\begin{multline}\label{eq:U|>}
  U |\cS_1 S^0_1 \dots \cS_k S^0_k\ket \\
    = (-)^{\sum_\tau(\cS_\tau+S^0_\tau)}
      |\cS_1 (-S^0_1) \dots \cS_k (-S^0_k)\ket.
\end{multline}
In particular, the state $|\cS_1 0 \dots \cS_k 0\ket$, where the shell
is half filled for all $\tau$, is an eigenstate of $U$ with eigenvalue
$(-1)^{\sum_\tau\cS_\tau}=(-1)^{(k\Omega-\sum_\tau v_{\tau})/2}$.

Now consider a Hamiltonian
\begin{multline}\label{eq:Ham}
  H = \sum_\tau \epsilon_\tau n_\tau \\[-5pt]
      + \sum_\text{$\begin{array}{c}\tau_1\tau_2\\[-3pt]
        \alpha_1\alpha_2\alpha'_1\alpha'_2\end{array}$}
      \frac1{2(1+\delta_{\tau_1\tau_2})} \bra\alpha_1\alpha_2|
        v_{\tau_1\tau_2} |\alpha'_1\alpha'_2\ket \\[-15pt]
      a_{\alpha_1\tau_1}^\dag a_{\alpha_2\tau_2}^\dag
      a_{\alpha'_2\tau_2} a_{\alpha'_1\tau_1},
\end{multline}
with a convention to be followed from now on that a variable denoted
by a Greek letter takes any value available to $m$ or $\mb$, where $m$
is the magnetic quantum number restricted to $m>0$, and $|\mb\ket$ is
the time-reversed of the state $|m\ket$. This choice of $|m\ket$ and
$|\mb\ket$ implies that $U$ commutes with the angular momentum
vector. The interaction matrix element is assumed symmetric in the
particles 1 and 2,
\begin{equation}\label{eq:symtau}
  \bra\alpha_1\alpha_2|v_{\tau_1\tau_2}|\alpha'_1\alpha'_2\ket
  = \bra\alpha_2\alpha_1|v_{\tau_2\tau_1}|\alpha'_2\alpha'_1\ket,
\end{equation} and antisymmetrized for
$\tau_1=\tau_2$, that is,
\begin{equation}\label{eq:antisym}
  \bra\alpha_1\alpha_2|v_{\tau\tau}|\alpha'_1\alpha'_2\ket
  = - \bra\alpha_1\alpha_2|v_{\tau\tau}|\alpha'_2\alpha'_1\ket,
\end{equation}
and the interaction is supposed to be rotationally invariant and
invariant under time reversal. The rotational invariance implies
\begin{equation}\label{eq:rotinv}
  \sum_\beta \bra\alpha\beta|v_{\tau_1\tau_2}|\alpha'\beta\ket
  = \delta_{\alpha\alpha'} 2\Omega \vb_{\tau_1\tau_2}
\end{equation}
with
\begin{equation}\label{eq:vbar}
  \vb_{\tau_1\tau_2} = \frac1{(2\Omega)^2}\sum_{\alpha\beta}
    \bra\alpha\beta|v_{\tau_1\tau_2}|\alpha\beta\ket,
\end{equation}
and by the time-reversal invariance we have
\begin{equation}\label{eq:timeinv}
  \bra\alpha_1\alpha_2|v_{\tau_1\tau_2}|\alpha'_1\alpha'_2\ket
    = \bra\alb'_1\alb'_2|v_{\tau_1\tau_2}|\alb_1\alb_2\ket
\end{equation}
with $|\alb\ket=|\mb\ket$ and $-|m\ket$ for $\alpha=m$ and $\mb$. A
simple calculation by means of \eqref{eq:conj} then shows that by
particle-hole conjugation,
\begin{multline}\label{eq:Ham}
  H \mapsto H + 2\Omega \sum_\tau \epsilon_\tau
    + \tfrac12 (2\Omega)^2 \sum_{\tau\tau'} \vb_{\tau\tau'} \\
    - \sum_\tau  \left(2\epsilon_\tau
      + 2\Omega \sum_{\tau'} \vb_{\tau\tau'}\right) n_\tau.
\end{multline}

In a space with definite $n_\tau$ the terms in this expression after
the first one are constant. Therefore the transformation $U$ maps each
eigenstate of $H$ in such a space to an eigenstate of $H$ in the image
of that space whose energy differs from the original one only by these
constant terms. Particularly in the midshell space with $S^0_\tau=0$
for all $\tau$, the Hamiltonian $H$ commutes with $U$ so that
$(-1)^{\sum_\tau\cS_\tau}=(-1)^{(k\Omega-\sum_\tau v_{\tau})/2}$ is
conserved. Because the factor $(-1)^{k\Omega}$ is a constant, this is
the conservation law observed by Escuderos, Zamick, and
Bayman~\cite{Es05}.

\section{\label{sec:vt}Conservation laws in $v$ and $t$}

Now assume, in the nuclear case with $k=2$, that $H$ is charge
invariant. I shall show how it then follows from the conservation law
just derived that in the subspace of a $j=7/2$ shell with
$n_n=n_p=\Omega=4$ and angular momentum and isospin $I=T=0$, the
$\Sp(2\Omega)$ irrep $vt=61$ is conserved.

\begin{figure*}
  {\center\includegraphics[width=\textwidth]{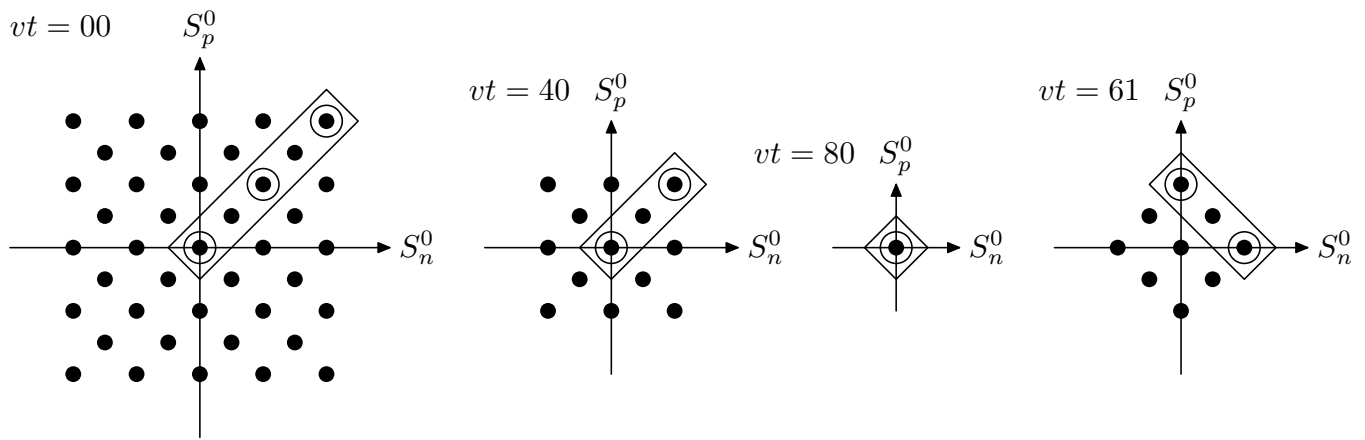}\par}
  \caption{\label{fig:48Crw}Weight diagrams of the $\Sp(4)$ irreps
    corresponding for $j=7/2$ to the $\Sp(2\Omega)=\Sp(8)$ irreps
    $vt=00$, $40$, $80$, and $61$. Rings indicate weights
    $(\cS_n,\cS_p)$ with integral coordinates.}
\end{figure*}

The $\Sp(2\Omega)$ irreps contained in this subspace are $vt=00$,
$40$, $61$, and $80$~\cite{Fl52}. The weight diagrams of the
corresponding $\Sp(4)$ irreps are shown in Fig.~\ref{fig:48Crw}.
Their ranges~\eqref{eq:Sp4>Sp2nxSp2p} are seen to lie on single lines.
In the case of $vt=61$, this line slopes downwards and
$\cS_n+\cS_p=\text{constant}=1$. For $t=0$, the line is $S^0_n=S^0_p$,
so $\cS_n+\cS_p=2\cS_n$. This increases from zero in steps of one, but
if $\cS_n$ is half-integral, the $\Sp(2)_n$ irrep has no state with
$S^0_n=0$, that is, $n_n=\Omega$. Therefore only integral $\cS_n$
contribute, and $\cS_n+\cS_p$ is even. Because the $\Sp(2\Omega)$
irreps $vt=00$, $40$, and $80$ thus have only even $\cS_n+\cS_p$, and
$vt=61$ only odd $\cS_n+\cS_p$, it follows that the latter is
conserved. The weights $(\cS_n,\cS_p)$ with integral coordinates are
indicated by rings in Fig.~\ref{fig:48Crw}.

When, more generally, can such a situation occur? First of all, by the
first condition \eqref{eq:weights}, for $(0,0)$ to be a weight of the
$\Sp(4)$ irrep, $(\lambda+\mu)/2$, and therefore $t=(\lambda-\mu)/2$,
must be integral. If $\cS_n+\cS_p$ is to have constant parity for
integral $\cS_n$ and $\cS_p$ within a given $\Sp(4)$ irrep, no
$(\cS_n,\cS_p)$ with integral $\cS_n$ and $\cS_p$ must have a neighbor
in the horizontal or vertical direction. This is satisfied for $t=0$,
that is, $\lambda=\mu$. The weight diagram then has the shape of the
upright square, possibly shrunk to a point, encountered above for
$vt=00$, $40$, and $80$, and the range \eqref{eq:Sp4>Sp2nxSp2p}
shrinks to the line segment from the center to the upper right corner
of this square. For integral $\cS_n$ and $\cS_p$, the sum
$\cS_n+\cS_p$ is even. If $t>0$ then $(t-1,1)$ is a possible value of
$(\cS_n,\cS_p)$. Then $(t,1)$ must not be a weight, which requires
$t=(\lambda+\mu)/2$, or $\mu=0$. Equivalently, $v/2+t=\Omega$. The
weight diagram then has the shape of the tilted square encountered
above for $vt=61$, and the range \eqref{eq:Sp4>Sp2nxSp2p} shrinks to
the upper right side of this square. In this case, $\cS_n+\cS_p=t$.

In the subspace of $\cV$ with $n_n=n_p=\Omega$, angular momentum $I$,
and isospin $T$, the states with a definite parity of $\cS_n+\cS_p$
thus carry definite irreps of $\Sp(2\Omega)$ when all such irreps
present have either $t=0$ or $v/2+t=\Omega$. Using Flowers's
classification of a basis for $\cV$ by (the unitary subgroups of) the
chain $\GL(2\Omega)\supset\Sp(2\Omega)\supset\SL(2)$ for
$j=3/2$--$7/2$, where $\GL(2\Omega)$ and $\SL(2)$ are the general and
special linear groups, and the infinitesimal algebra of the latter is
the algebra of angular momentum~\cite{Fl52}, one can list for these
$j$ all such subspaces, along with the classification of a basis by
$v$ and $t$. This is done in Tables~\ref{tab:3/2&5/2}
\begin{table}
  \caption{\label{tab:3/2&5/2}For each subspace of $\cV$ with
    $n_n=n_p=\Omega$ and given $I$ and $T$ containing only states from
    $\Sp(2\Omega)$ irreps with $t=0$ or $v/2+t=\Omega$ the
    classification of a basis by such irreps is indicated. Irreps
    with even and odd $\cS_n+\cS_p$ are listed in separate columns.
    $j=3/2$ and $5/2$.}
\begin{ruledtabular}
\begin{tabular}{ccccc}
$j$&$I$&$T$&$vt$ with even $\cS_n+\cS_p$&$vt$ with odd $\cS_n+\cS_p$\\
\hline
3/2&0&0&00&\\
&0&2&00&\\
&1&1&20&\\
&2&0&40&21\\
&2&1&&21\\
&3&1&20&\\
&4&0&40&\\
5/2&0&1&00 40&\\
&0&3&00&\\
&1&0&20 60&41\\
&1&1&&41\\
&1&2&20&\\
&2&0&&41\\
&3&0&20 60$^2$&41$^2$\\
&3&1&40&41$^2$\\
&3&2&20&\\
&4&0&60&41\\
&5&0&20 60&41\\
&5&1&40&41\\
&5&2&20&\\
&6&0&60&41\\
&6&1&40$^2$&41\\
&7&0&60&41\\
&7&1&&41\\
&8&1&40&\\
&9&0&60&
\end{tabular}
\end{ruledtabular}
\end{table}
and~\ref{tab:7/2}.
\begin{table}
  \caption{\label{tab:7/2}As Table~\ref{tab:3/2&5/2} for $j=7/2$.}
\begin{ruledtabular}
\begin{tabular}{ccccc}
$j$&$I$&$T$&$vt$ with even $\cS_n+\cS_p$&$vt$ with odd $\cS_n+\cS_p$\\
\hline
7/2&0&0&00 40$^2$ 80&61$^2$\\
&0&1&&61$^2$\\
&0&2&00 40$^2$&\\
&0&4&00&\\
&1&0&80&61$^2$\\
&1&3&20&\\
&3&0&40$^2$ 80$^2$&61$^5$\\
&3&3&20&\\
&5&0&40$^2$ 42 80$^3$&61$^6$\\
&5&3&20&\\
&7&0&40$^2$ 80$^3$&61$^6$\\
&7&3&20&\\
&8&0&40$^3$ 42 80$^4$&61$^6$\\
&9&0&40 80$^3$&61$^4$\\
&10&0&40$^2$ 80$^3$&61$^4$\\
&11&0&80$^2$&61$^2$\\
&12&0&40 80$^2$&61$^2$\\
&12&1&60$^2$&61$^2$\\
&12&2&40&\\
&13&0&80&61\\
&13&1&60$^2$&61\\
&14&0&80&61\\
&14&1&&61\\
&15&1&60&\\
&16&0&80&
\end{tabular}
\end{ruledtabular}
\end{table}
The Hamiltonian does not mix the states in the last but one and last
columns. When $T>0$, this is true not only for $n_n=n_p$, that is,
$T^0=0$, but holds in the entire isospin multiplet (which has
$n=2\Omega$) because $\Sp(2\Omega)$ commutes with $\vT$. The reader
may check that for each $I$ and $T$ in Table~\ref{tab:7/2}, the
dimensions of the subspaces with even and odd $\cS_n+\cS_p$, that is,
even and odd $(v_n+v_p)/2$, concur with those found numerically by
Escuderos, Zamick, and Bayman~\cite{Es05}.

For $j=3/2$, all $\Sp(2\Omega)$ irreps present for $n=2\Omega=4$ have
either $t=0$ or $v/2+t=\Omega$. It is seen from
Table~\ref{tab:3/2&5/2} that all eigenstates of $H$ then have definite
$v$ and $t$. This is known already to hold for other reasons for any
$n$~\cite{Ra52}. For $j=5/2$, the only $\Sp(2\Omega)$ irrep present
for $n=2\Omega=6$ other than those in Table~\ref{tab:3/2&5/2} is
$vt=21$. For $j=7/2$, the only $\Sp(2\Omega)$ irreps present for
$n=2\Omega=8$ other than those in Table~\ref{tab:7/2} are $vt=21$ and
41.

\section{\label{sec:sum}Summary}

The irreps of the symplectic group $\Sp(4)$, which is the commutator
group of the group $\Sp(2\Omega)=\Sp(2j+1)$ of symplectic
transformations of the single-nucleon states in the space $\cV$ of any
numbers of nucleons of both kinds in a $j$ shell~\cite{He60}, were
shown to be the direct sum of different single irreps of
$\Sp(2)_n\times\Sp(2)_p$, where $\Sp(2)_n$ and $\Sp(2)_p$ are Kerman's
quasispin groups~\cite{Ke61} for neutrons and protons. A rule for the
range of $\Sp(2)_n\times\Sp(2)_p$ irreps in this decomposition is
given in~\eqref{eq:Sp4>Sp2nxSp2p} and its surrounding text. The theory
of particle-hole conjugations was reviewed, and conservation of the
parity of $(v_n+v_p)/2$ in the system with $\Omega$ valence nucleons
of both kinds, where $v_n$ and $v_p$ are the neutron and proton
seniorities~\cite{Ra43,Ra49}, derived in a simple manner. This
derivation only requires that the Hamiltonian is the sum of a one- and
a two-body term and invariant under rotations and time reversal. The
conservation of the parity of $(v_n+v_p)/2$ was observed by Escuderos,
Zamick, and Bayman in calculations for $^{48}$Cr in the $1f_{7/2}$
shell model~\cite{Es05}. Combining these results, I explained the
conservation of the $\Sp(2\Omega)$ irrep $(v,t)=(6,1)$, where $v$ is
the seniority, and $t$ the reduced isospin~\cite{Fl52}, for angular
momentum and isospin $I=T=0$ found in my own calculations of this
kind~\cite{Ne14}. Using Flowers's classification of a basis for $\cV$
by the chain $\GL(2\Omega)\supset\Sp(2\Omega)\supset\SL(2)$ for
$j=3/2$--$7/2$~\cite{Fl52}, I listed for these $j$ all the cases of
angular momentum $I$ and isospin $T$ where an analogous mechanism
takes effect, along with the $\Sp(2\Omega)$ classification of a basis
in each case. For $j=7/2$, this list explains in each such case the
dimensions of the subspaces with definite parities of $(v_n+v_p)/2$
observed in the calculations by Escuderos, Zamick, and Bayman.

\begin{acknowledgments}
  I am indebted to Larry Zamick for suggesting to me that the
  numerically based observations in \cite{Es05} and \cite{Ne14} might
  have related explanations.
\end{acknowledgments}

\bibliography{cons}

\end{document}